\documentclass{article}
\usepackage{amsmath}
\usepackage{amssymb}
\usepackage{bm}

\usepackage{graphicx}
\usepackage{dcolumn}
\usepackage{bm}
\usepackage{color}
\usepackage{float}
\usepackage{layout}

\usepackage{caption}
\newcommand{\dif}{{\rm d}}

\newcommand{\etal}{{\it et al}}
\definecolor{lightgray}{gray}{0.5}
\begin{document}
\title{Strengthening our grip on food security by encoding physics into AI.}
\author{Marcel B.J. Meinders$^{1,2,*}$, Jack Yang$^{1,3}$, and Erik van der Linden$^{1,3}$}
%
\date{
{\it
$^1$ Wageningen University and Research Centre, Wageningen, The Netherlands.\\
$^2$ Wageningen Food and Biobased Research, Wageningen, The Netherlands.\\
$^3$ Wageningen University, Wageningen, The Netherlands.\\
$*$ Corresponding author marcel.meinders@wur.nl\\
{\center\bf Confidential}
}
\today}

\maketitle
\graphicspath{ {/Figures/} }

\section*{Abstract}
Climate change will jeopardize food security. Food security involves the robustness of the global agri-food system. This agri-food system is intricately connected to systems centering around health, economy, social-cultural diversity, and global political stability. A systematic way to determine acceptable interventions in the global agri-food systems involves analyses at different spatial and temporal scales. Such multi-scale analyses are common within physics. Unfortunately, physics alone is not sufficient. Machine learning techniques may aid. We focus on neural networks (NN) into which physics-based information is encoded (PeNN) and apply it to a sub-problem within the agri-food system. We show that the mean squared error of the PeNN is always smaller than that of the NNs, in the order of a factor of thousand. Furthermore, the PeNNs capture extra and interpolation very well, contrary to the NNs. It is shown that PeNNs need a much smaller data set size than the NNs to achieve a similar mse. Our results suggest that the incorporation of physics into neural networks architectures yields promise for addressing food security.

\section{Introduction}
It is generally accepted that climate change will jeopardize food security in the near future. The robustness of the agri-food system is intricately connected to the systems that revolve around health, economy, social-cultural diversity, and global political stability, to name but a few. All these systems are continuously changing in an interdependent, self-organizing, and adaptive manner, have various levels of detail and require various disciplines for their description and understanding. A systematic way to determine acceptable interventions to have these systems continuously adapt together towards a desired level of food security involves multi-scale analyses and integration of different disciplines, while simultaneously incorporating the adaptive dynamics at all relevant levels. This has been coined before as a complex systems approach. An according systematic method was proposed for determining optimal intervention strategies to steering complex (agri-food) systems~\cite{van2014complex}. The approach was argued to thrive from integrating various methodologies, including scaling, multi-scale modeling, machine learning, including graphical and evolutionary algorithms (\cite{van2014complex} and specific references therein). Additionally, computational techniques used in a complex system approach have been reviewed in another article, by Perrot et al.~\cite{perrot2016some}. However, in both articles, no specific methodology was articulated on how to specifically integrate multi-scale modeling and machine learning techniques. To this end, the current article addresses a way forward, and illustrates this with an example problem in a small sub-set of the agri-food system.

For our purpose, we pick a material science area within agri-food. As we know, multi-scale analyses have been an approach in this physics-oriented material science area for some time. The specific multiple spatial and temporal scales for this type of problem range from the molecular to macroscopic length scale, and from femtosecond to years, respectively. 

Describing relationships between properties on a molecular to macroscopic scale is a complex endeavor since multiple connections exist between different scales. Furthermore, such properties evolve, and the according dynamics usually cannot be covered by formal analytic descriptions. This according complexity makes the use of available mechanistic models alone difficult. A common strategy to relate molecular to macroscopic properties is to use an intermediate length scale, known as a meso- or microstructural length scale, and describe the macroscopic properties in terms of properties of this length scale. This strategy has its challenges, attributed to the ill-reliability of microscale models, the difficulty in simulating the micro-scale properties accurately, and the often intricate “entanglement” between micro-structural and macroscopic scale properties~\cite{weinan2020integrating}.

In order to accommodate these challenges described in the previous paragraph, machine learning (ML) can be an asset. ML models are able to find structures and patterns in data sets and thus able to improve mechanistic physics multi-scale models. Reversely, physical information may aid in structuring the ML algorithms and make these more efficient, i.e. using less data for a given prediction accuracy. We note that the number of data used for a sufficient accuracy of ML is often much larger than practically available. This is where physics information can come to the rescue. From that perspective, “aiding” ML models by means of mechanistic models seems a promising way forward. More generally, combining physics-based multi-scale modeling with ML techniques may cleverly solve two problems at the same time. Physics may provide the structuring of ML techniques, and turning correlative relations into casual ones. Attempts to combine ML with physics-based modeling for dynamics of lake temperature~\cite{jia2021physics,daw2017physics,read2019process}, and phosphorus concentration~\cite{hanson2020predicting} have shown that one can obtain better predictions with a smaller number of data and for scenarios that are distinct from the training scenario used in the ML algorithm itself. A recent review on physics-informed machine learning can be found in ~\cite{faroughi2022physics} and with application focus on life sciences in~\cite{alber2019integrating}. In these reviews, the challenges and possible routes forward to model spatial-temporal evolution combining physics and ML are clearly addressed. To capture the spatio-temporal dynamics, one can use partial differential equations to express physics-based conservation laws, where such conservation laws can be constructed from constitutive laws that represent the local behaviour, in combination with using ordinary differential equations and their spatial derivatives. This has been illustrated in more detail for rheology-informed neural networks by Mahmoudabadbozchelou and Jamali~\cite{mahmoudabadbozchelou2021rheology}. Interestingly, Sadaat \etal~\cite{saadat2022data} report the use of a platform of possible constitutive models to have the NN pick from in order to optimize its predictions for more complex fluid behavior. These constitutive models need not representing the entire physics of the problem in all its details. Interestingly, parts of constitutive models can be captured by means of applying scattering under flow, as demonstrated by Young \etal  while using Small Angel X-ray scattering on dilute rod suspensions~\cite{young2023scattering}. In trying to take into account spatio-temporal non-linear features, Dabiri \etal~\cite{dabiri2023fractional} report the use of fractional derivatives to incorporate into the NN models. Fractional derivatives are known to represent the presence of memory effects, which may be uncovered by introducing hidden variables that describe local effects, as addressed in Weinan \etal~\cite{weinan2020integrating}. In that work, an exciting example of introducing temporal (dynamical) information has been addressed in the form of so-called “recurrent neural networks”. The neural networks are machine learning models for time series. These models use hidden variables, making the relationships, as expressed in the models, local.  If no hidden variables are being used, one effectively introduces memory effects~\cite{weinan2020integrating}.

There are several ways at our disposal to add physical information to a neural network. For example, one can ascribe physics information to nodes in the network~\cite{willard2020}. Alternatively, one can add physical information regarding symmetries that need to be obeyed. Another option could be the use of physics-based model data as input to AI models. For a survey on recent progress in various fields, the reader is referred to Willard \etal~\cite{willard2020}. It is noted that humans can develop physics-based architectures of neural networks, but this can be automated as well (\cite{willard2020} references 13,73, 115).  A concrete set of examples of improvement of neural network performance, which at the same time preserves the correctness of the physics, has been recently published by Takeishi and Kalousis~\cite{takeishi2021}.

In regards to the agri-food area, in particular in applying ideas on combining multi-scale modeling with ML directly, i.e. without the need to reprogram the neural networks ML part, a review of Peng \etal~\cite{peng2020multiscale} is worthwhile to mention.  Works with more direct embedding in food science that address the combination of physics information and ML, without reprogramming the neural networks that underlie the ML are, for example, found elsewhere~\cite{lie2023machine}. In this same area already some reviews can be found ~\cite{bhagya2022comprehensive,datta2022computer}. 

In the current article, we quantify the effects of including physics information in the architecture of NNs. We investigate uncertainty in prediction as a function of training set size, and effects on uncertainty/errors in inter- and extrapolating beyond a training set. We specifically look into a relatively simple problem of protein stabilized oil droplets aggregating into clusters, and how the cluster size distribution in turn will determine the shear viscosity versus shear strain. The problem becomes non-linear as the flow influences the aggregation and vice versa. The problem becomes more complex as one adds more experimental parameters that are known to influence the viscosity, such as the pH, protein type etc. The example lends itself to demonstrate the improvement by integrating physics into a NN.

\section{Flow of complex food fluids}
Many complex foods are in fact dispersions, which are composed of fluid or solid particles dispersed in a fluid. Viscosity is an important parameter because it controls texture, consumer acceptance and processing conditions. Amusingly, this also holds for non-food materials like paints. Due to the complex nature of the dispersions, the viscosity depends on the rate of deformation. For our purpose, we focus here on shear deformation. 

The shear rate-dependent viscosity depends on the structure of the dispersions and the interactions between the dispersed particles. The interactions control the assembly of (primary) particles into clusters. During the deformation of the complex fluid, these clusters can break up or can be formed, depending on time, shear rate, particle concentrations, cluster sizes, and inter-particle and inter-cluster interactions and temperature. Many studies have been published to describe the shear-rate dependent viscosity of complex dispersions~\cite{derkach2009rheology, rao2014rheology, larson2019,mcclements2023modeling}. Models to describe these kind of shear rate-dependent viscosity should contain the structural dynamics. One such model is the constitutive model by Quemada and coworkers~\cite{quemada1977,quemada1998a,berli2000rheological,berli2002modeling, quemada2002energy}, which describes the rheology of complex colloidal systems in a large range of volume fractions using key physical parameters. Because the model fits very well with viscosities of food dispersions~\cite{derkach2009rheology, jansen2001viscosity, mcclements2023modeling, rao2014rheology} (see also figure~\ref{fig:mayo}), we used the Quemada model to study the predictive power of different NNs, and hybrid neural networks containing physics, for predicting the shear rate dependent viscosity of complex dispersions. 

\subsection{Quemada model}
\begin{figure}[htbp]
    \centering
        \includegraphics[clip,width=0.4\textwidth]{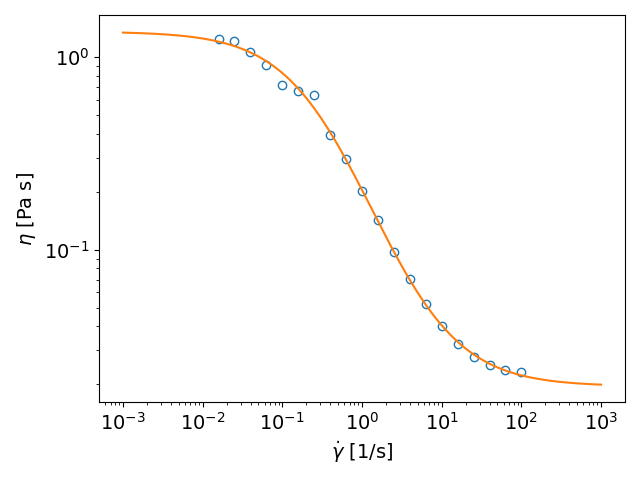}
    \caption{Example of a flow curve of an oil-in-water emulsion and Quemada-model-fit.}
    \label{fig:mayo}
\end{figure}

The Quemada model describes the viscosity of complex fluids in terms of an effective volume fraction, which depends on the properties of the primary particles, including diffusion coefficients, interaction parameters, packing fraction of particles in a cluster, and shear rate. Here, we will focus on the semi-stationary regime, where the system is in equilibrium at a certain shear rate. The starting point of the model is the relation between the viscosity of the fluid $\eta$ as function of the effective volume fraction $\phi_e$. This is given by
\begin{equation}
\eta = \eta_f \left(1-\frac{\phi_e}{\phi_m} \right)^{-2}
\end{equation}
with $\eta_f$ the viscosity of the continuous phase, and $\phi_m$ the maximum volume fraction. 
The effective volume fraction is given by
\begin{equation}
\phi_e = \phi_{pA}/\varphi + (\phi_p-\phi_{pA}) = \phi_p \left( 1 + CS\right) \label{eq:phie}
\end{equation}
with $\phi_p$ the volume fraction of the primary particles, $C=1/\varphi+1$, a compactness factor where $\varphi$ is the volume of the particles in a cluster divided by the volume of that cluster. $S$ is a structural parameter defined as the ratio between the volume fraction of primary particles in the cluster $\phi_{pA}$ and the total volume fraction of primary particles $\phi_p$
\begin{equation}
S = \frac{\phi_{pA}}{\phi_p} 
\end{equation}
The structural parameter follows a certain kinetic reaction scheme, which, in its basic form, reads 
\begin{equation}
\frac{\dif S}{\dif t} = \kappa_D (S_0-S) - \kappa_h (S-S_\infty) 
\end{equation}
with $\kappa_D$ and $\kappa_h$ characteristic relaxation rates of Brownian (diffusion) and hydrodynamic (shear stress) forces. $S_0$ and $S_\infty$ correspond to the value of the structural parameter at zero and infinite shear rate. Additional terms related to particle interactions can be added~\cite{quemada2002energy}. The steady-state value of the structural parameter at a certain shear-rate $\dot\gamma$ is then given by
\begin{equation}
S = \frac{S_0 + \theta S_\infty}{1+\theta} \label{eq:s}
\end{equation}
with
\begin{equation}
\theta = \frac{\kappa_h}{\kappa_D}=\frac{\dot\gamma}{a^2/D_p}=Pe=\frac{6\pi\eta_fa^3\dot\gamma}{k_BT} \label{eq:theta}
\end{equation}
and $a$ is the size of the primary particle, $D_p$ the diffusion coefficient of the primary particles, $k_B$ Boltzmann's constant, $T$ temperature, and $Pe$ the P{\'e}clet number.
The above equations give the fluid's viscosity in the stationary state as a function of the key parameters
\begin{equation}
\eta = \eta(\dot{\gamma}, \phi_p, S_0, S_\infty, \eta_f, a, T) \label{eq:eta}
\end{equation}
Figure~\ref{fig:mayo} shows an example of a measured flow curve of a pea protein stabilized oil-in-water emulsion, with an oil volume fraction of 0.5. The figure also shows the fit with the Quemada model with $\dot{\gamma}_c = \frac{6\pi\eta_fa^3}{k_BT}=4.7$ (corresponding to an average droplet size of $a=1$~$\mu$m), $CS_0=0.25$ (indicating that less than about 25\% of the oil droplets are flocculated at zero shear rate) and $CS_\infty=0$ (indicating that all clusters broke up at a high shear rate). 

\subsection{Neural Networks and Physics-encoded Neural Networks}
To investigate the predictive power NNs and PeNNs, shear rate dependent viscosity data was generated with the Quemada model and used to train and validate NNs and PeNNs with different architectures. In order to align with physical experiments, we aim to predict $y=\eta$ from a set of input parameters $\{x_i\}$. In an experimental setting, $\eta$ is measured as a function of the shear rate $\dot{\gamma}$, so $\dot{\gamma}\in \{x_i\}$. Various other input characteristics of the dispersion can be measured or are known. For example, when preparing a sample, the volume fraction $\phi_p$ of the primary particles is known, like the amount of oil in an emulsion. Other parameters can be measured and estimated using various different, often indirect, techniques, like e.g. the particle and cluster size distributions from light scattering techniques and inter-particle forces from DLVO theory. 

The NNs and PeNNs were built using Keras in combination with TensorFlow. 
Data was generated and used to train the NNs and PeNNs with different architectures. Because the shear rate and viscosity can vary over several orders of magnitude, their logarithm was used to train the NNs and PeNNs. As the other parameters are in the order of 1, no other scaling was applied.

We investigated two cases to compare. In case one, we predict the viscosity as a function of two input parameters, while in case two, we predict the viscosity as a function of four input parameters. 

\subsubsection{Case 1}
For case one, two different NNs were studied. One with an architecture consisting of two hidden dense layers. The first layer has $n_1=32$ neurons and is densely connected to the $n_{in}=2$ input nodes, while the second layer has $n_2=8$ neurons. Another NN studied has an architecture consisting of three hidden dense layers. The first, second and third layer has $n_1=128$, $n_2=32$ and $n_3=8$ neurons, respectively. For both NNs, the first layer is densely connected to the $n_{in}=2$ input nodes, while the last layer is densely connected with the output layer having $n_{out}=1$ neuron. The rectified linear (ReLu) activation function was used for the hidden layers, while a linear activation function was used for the output layer. The number of trainable parameters of the 2-32-8-1 and 2-128-32-8-1 networks are 371 and 4787, respectively. 

The PeNN consist of 3 layers, each corresponding to a physical quantity, in other words, the activation functions are completely physics-based. In this sense, this PeNN is actually totally dominated by physics, to illustrate the importance of physics in a NN. The first layer corresponds to the structure parameter $S$. It has one input $x=\log\dot\gamma$ and activation function $S_{act}= 1/(10^xw + 1)$. Here, $w$ is a trainable parameter and should correspond to $\frac{6\pi\eta_fa^3}{k_BT}$ as can be simply derived from equations~\ref{eq:s} and \ref{eq:theta}. The second layer corresponds to the effective volume fraction $phi_e$ having two inputs, being the output of the S-layer and the input $phi_p$. The activation of the $phi_e$-layer is $\phi_{e,act}= x_1(x_2w + 1)$, with input $\bm{x} = [x_1, x_2]$ and $w$ a trainable parameter corresponding to $C$ (equation~\ref{eq:phie}). The third layer corresponds to the viscosity with activation function $eta_{act} = -2\log(1-xw)+b$, where $w$ and $b$ are trainable parameters, corresponding to $1/phi_m$ and $\log\eta_f$, respectively, and $x$ is the input equal to the output of the effective-volume-fraction-layer. 

\begin{figure}[htbp]
    \centering
        \includegraphics[clip, trim=0cm 5.5cm 1cm 5.5cm, width=1.00\textwidth]{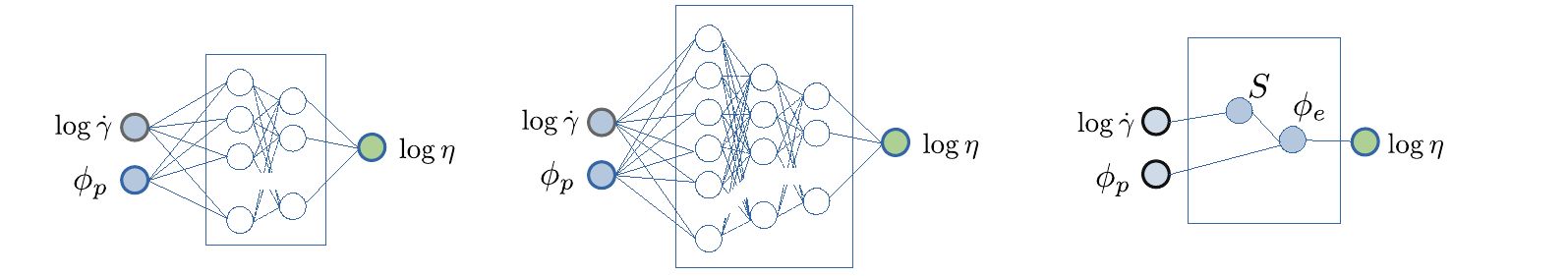}
    \caption{Schematic architecture of the NN (left, middle) and PeNN (right) configuration.}
    \label{fig:architect0}
\end{figure}

In case one, only the shear rate $\dot{\gamma}$ and particle volume fraction $\phi_p$ were varied and taken as input parameters for the constitutive model to generate the steady-state fluid viscosity $\eta$. Other parameters were taken constant, being $T=293$~K, $a=5$~nm, $C=2$, $S_0=1$, $S_\infty=0$, and $\eta_f=10^{-3}$~Pa.s. 

\subsubsection{Case 2}
\begin{figure}[htbp]
    \centering
        \includegraphics[clip, trim=0cm 4.5cm 1cm 5.5cm, width=1.00\textwidth]{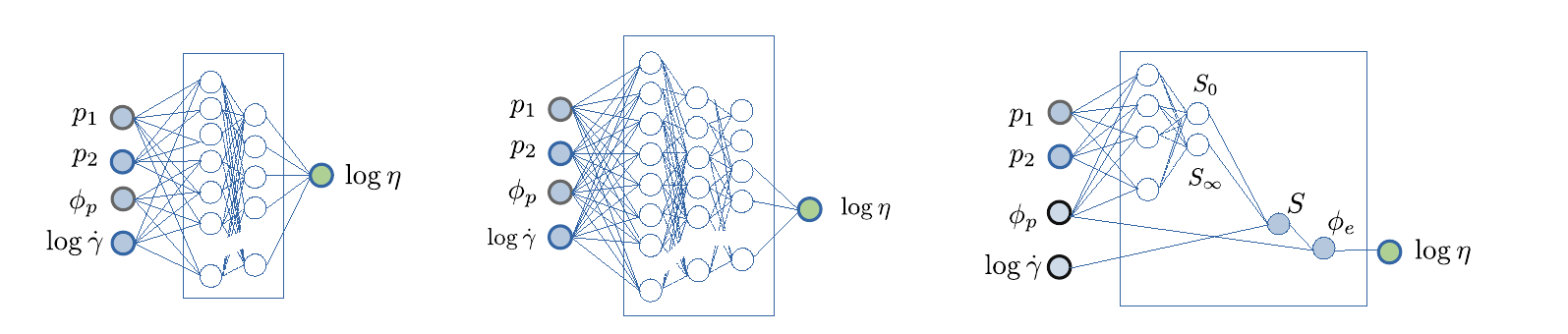}
    \caption{Schematic architecture of the NN (left, middle) and PeNN (right) configuration.}
    \label{fig:architect0_1}
\end{figure}

In case 2, also $S_0$ and $S_\infty$ were varied to generate flow curve data to train NNs and PeNNs with architectures, as depicted in figure~\ref{fig:architect0_1}. These parameters correspond to the structure parameter at zero and infinite shear rate, respectively. In this sense, this PeNN is not totally dominated by physics. In general, parameters are difficult to asses as they can be related to handling history, inter-particle forces, amount of protein denaturation, pH, salt concentration, etc. Here, we took two representative input parameters $p_1$ and $p_2$ and scaled them between 0 and 1. 

Similar as above, two different NNs were studied architectures consisting of two and three dense hidden layers consisting of 32-8 and 128-32-8 neurons, respectively hidden dense layers. For both, the first layer is densely connected to the $n_{in}=4$ input nodes, while the last layer is densely connected with the output layer having $n_{out}=1$ neuron. The rectified linear (ReLu) activation function was used for the hidden layers, while a linear activation function was used for the output layer. The number of trainable parameters of the 4-32-8-1 and 4-128-32-8-1 networks are 440 and 5048, respectively. 

The PeNN consist of 6 layers, of which the first 3 are a dense connected NN with outputs that should mimic the structure factor at low and high shear rate $S_0$ and $S_\infty$, respectively. The following and last 3 layers correspond each to the physical quantities, as explained above. The only exception is that the S-layer now has three inputs $\bm{x} = [x_1 x_2 x_3]$ corresponding to $[S_0 S_\infty \log\dot\gamma$] and physics-based activation function $S_{act}= (x_1 +10^{x_3}wx_2)/(10^{x_3}w + 1)$. Here, again $w$ is a trainable parameter and should correspond to $\frac{6\pi\eta_fa^3}{k_BT}$. The last layers are the same as described above.

\subsubsection{Data sets}
For both cases 1 and 2, various data sets were generated and used to test the performance of the NNs and PeNNs, varying in number of input parameters (as discussed above) as well as varying in number of data points per input parameter. The generated data was split into a training set (75\%, randomly chosen) used to train NNs and PeNNs to test the performance after training. During training, 10\% of the set was used for validation. The mean square error loss function was used as well as the Adam optimization algorithm with a learning rate of 1e-3 and decay of 5e-6. Training was stopped when, for 500 epochs, the value of the loss function did not show a decrease.

It is noted that in experiments, the number of data points for the shear rate $\dot\gamma$ is much larger than that for a parameter like the volume fraction of the primary particles $\phi_p$. This is because it is rather simple to obtain 100 or more data points ($\eta(\dot\gamma )$) for just one sample (with a certain $\phi_p$) in a viscosity measurement. Therefore, we also generated the data sets in a similar way: for each of $n$ different $\phi_p$'s (with $\phi_p \in \{\phi_1 \ldots \phi_n\}$), $N$ different $\log\dot\gamma$ (with $\dot\gamma \in \{\dot\gamma_1 \ldots \dot\gamma_N\}$ were chosen as input parameters, with $N>>n$. This yields $n\times N$ data triplets ($\eta(\phi_p,\dot\gamma )$. This set was randomly split into a training set (75\%) and a test set (25\%).  In general, the test set is used to check the performance of a NN to unseen data. However, although the NN did not see the data triplets ($\eta(\phi_p,\dot\gamma )$ of the test set, it did see numerous data with $\phi_p\in \{\phi_1 \ldots \phi_n\}$. In order to check if the NNs and PeNNs can also generalize to unseen volume fractions, thus how they  perform for $\phi_p \notin \{\phi_1 \ldots \phi_n\}$, a validation set was created by choosing randomly $n_{val}=1e4$ input parameters $(\phi_p,\log\dot\gamma )$ with $\phi_p$ between $\phi_{p,min}=0.01$ and $\phi_{p,max}=0.2$ and $\log\dot\gamma$ between $\log\dot\gamma_{min}=-5$ and $\log\dot\gamma_{max}=3$. 

Performance of the NNs and PeNNs was assessed by comparison of predicted and ground truth viscosity values and the calculation of the $R^2$-score and root mean square error. 

\section{Results and discussion}

\subsubsection{Case 1}
\begin{figure}[htbp]
    \centering
        \includegraphics[clip, trim=1cm 1.cm 1cm 1.cm, width=1.00\textwidth]{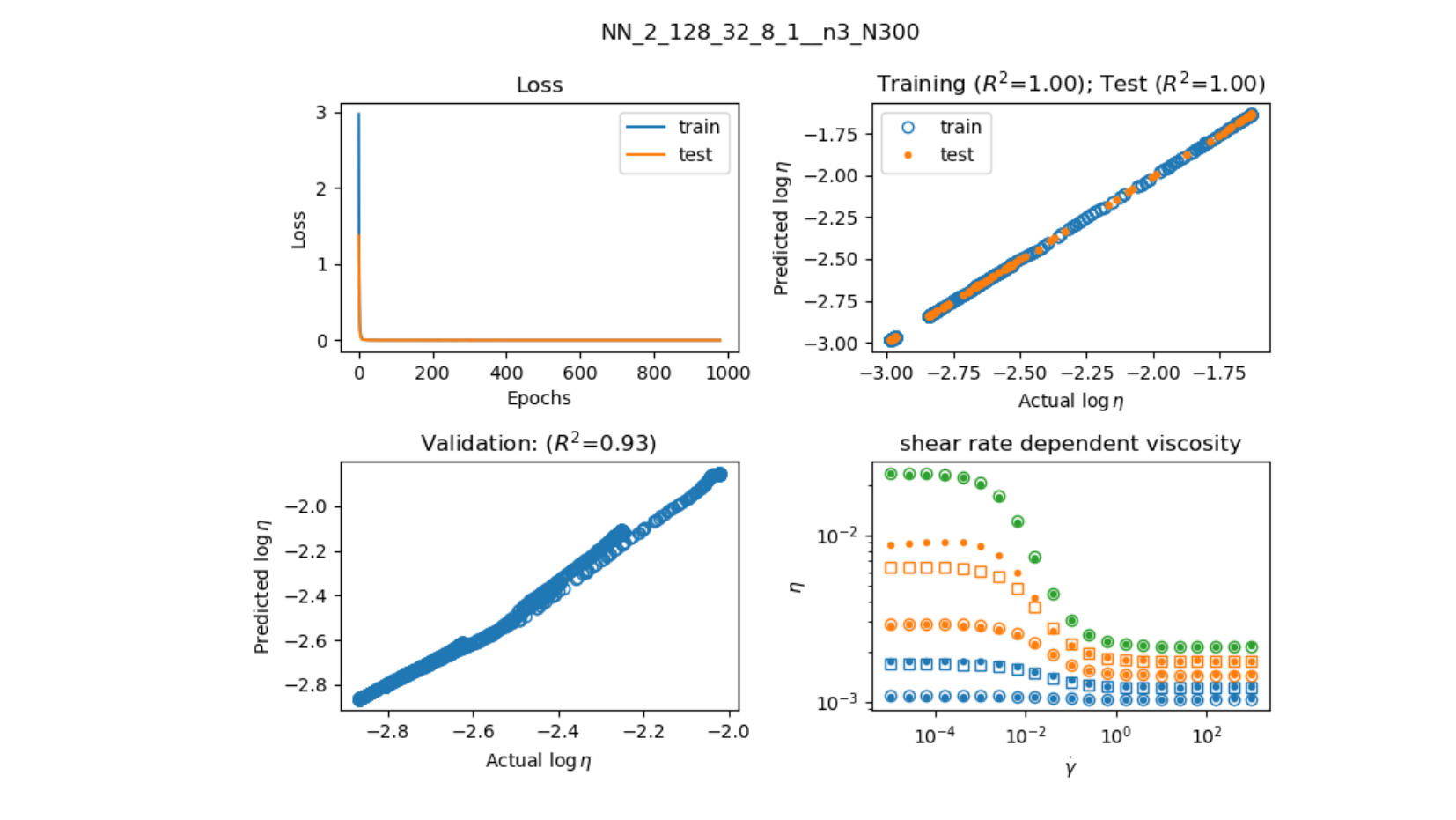}
    \caption{Results of the NN 2-128-32-8-1 neural network (see figure~\ref{fig:architect0_1}). This NN was trained and tested using a data set generated from $n=3$ different $\phi_p$ and for each $\phi_p$ $N=300$ different $\dot\gamma$. Top-left: loss as a function of number of epochs; Top-right predicted versus actual values of the training (blue) and test (orange) set; Bottom-left: predicted versus ground truth of the validation set containing unseen  $\phi_p$; Bottom-right: examples of ground truth (actual) shear rate dependent viscosity ($\circ$ (seen $\phi_p$) and $\square$ (unseen $\phi_p$)) and predicted shear rate dependent viscosity ($\bullet$). The colors indicate different $\phi_p$}
    \label{fig:1}
\end{figure}

Figure~\ref{fig:1} shows an example of the results of a NN for case 1, with two input nodes ($\phi_p$ and $\log\dot\gamma$), three hidden layers with 128, 32, and 8 neurons, respectively, and one output layer $\log\eta$. This NN was trained and tested using a data set generated from $n=3$ different $\phi_p$ and for each $\phi_p$ $N=300$ different $\dot\gamma$. The top-left panel shows the loss function as a function of the number of epochs. The top-right panel shows the predicted output as a function of the actual (ground truth) output, for the training set as well as the test set. 
The predicted values and actual values of the validation set, also containing $\phi_p$-values not seen by the NN, are plotted against each other in the left-bottom panel. 
In addition, the right-bottom panel of the figure shows the predicted and ground truth viscosity as function of shear rate and for different $\phi_p$, seen, as well as unseen during training.

\begin{figure}[htbp]
    \centering
        \includegraphics[clip, trim=1cm 0.5cm 1cm 1.2cm, width=1.00\textwidth]{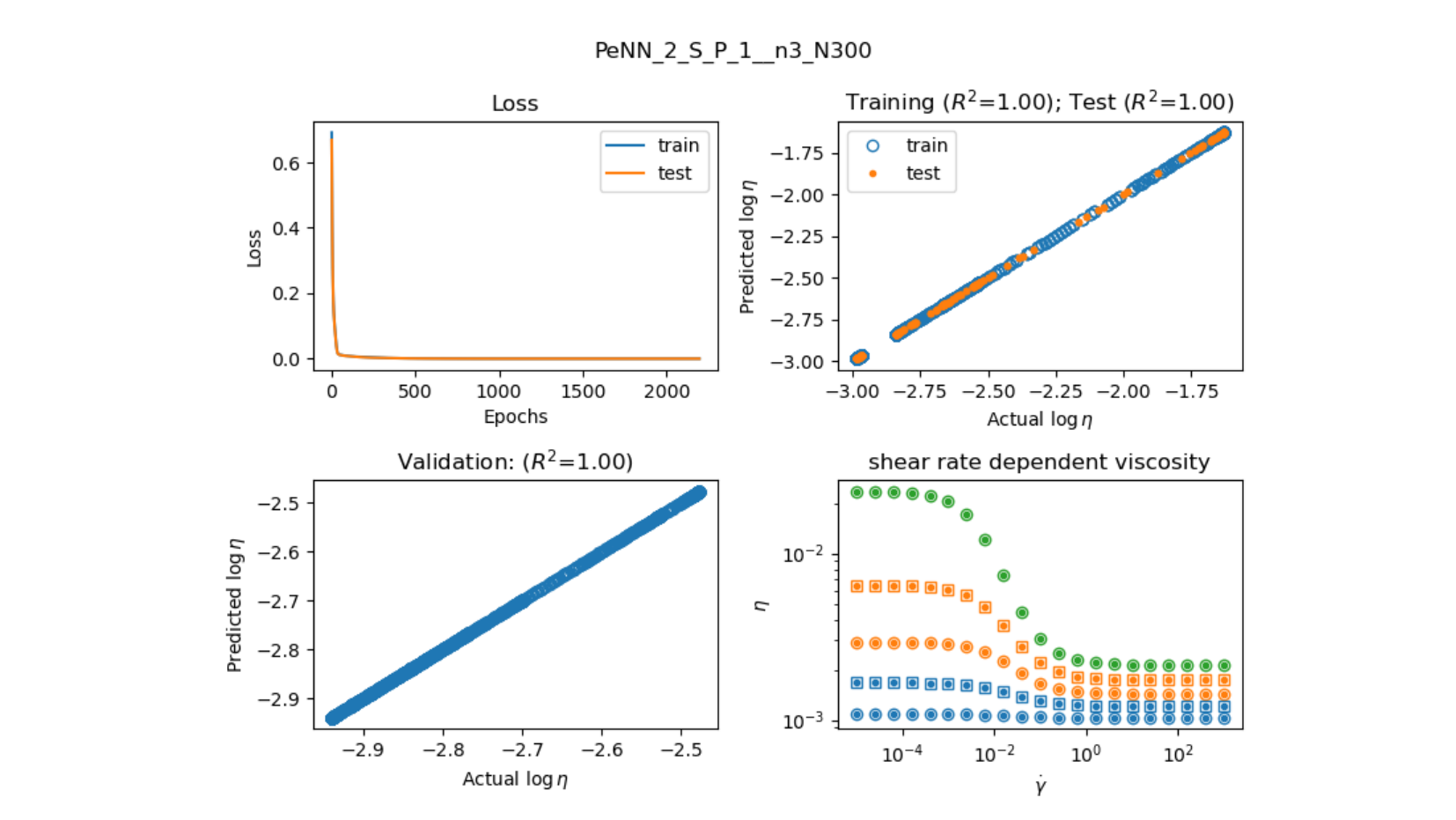}
    \caption{Results of the PeNN 2-1-1-1 [physical-encoded neural network (see figure~\ref{fig:architect0_1}). This NN was trained and tested using a data set generated from $n=3$ different $\phi_p$ and for each $\phi_p$ $N=300$ different $\dot\gamma$. Top-left: loss as a function of number of epochs; Top-right predicted versus actual values of the training (blue) and test (orange) set; Bottom-left: predicted versus actual values of the validation set containing unseen  $\phi_p$; Bottom-right: examples of actual shear rate dependent viscosity ($\circ$ (seen $\phi_p$) and $\square$ (unseen $\phi_p$)) and predicted shear rate dependent viscosity ($\bullet$). The colors indicate different $\phi_p$}
    \label{fig:2}
\end{figure}

Figure~\ref{fig:2} shows an example of the results of the PeNN for case 1, with two input nodes ($\phi_p$ and $\log\dot\gamma$), two layers completely physics-encoded, and one output layer $\log\eta$. The top-left panel shows the loss function as a function of the number of epochs. The top-right panel shows the predicted output as a function of the actual (ground truth) output, for the training set as well as the test set. The predicted values and actual values of the validation set containing $\phi_p$'s not seen by the PeNN are plotted against each other in the left-bottom panel. 
In addition, the right-bottom panel of the figure shows the predicted and ground truth (actual) viscosity as function of shear rate and for different $\phi_p$, seen, as well as unseen during training.

We conclude that in this case 1, replacing neural nodes by physics does improve NN significantly. The viscosity as function of shear rate and primary particle volume fraction are predicted very well, also for values of input parameters which the model has not been trained for. Thus interpolation and extrapolation are captured well for the PeNNs.

\subsubsection{Case 2}
\begin{figure}[htbp]
    \centering
        \includegraphics[clip, trim=1cm 0.5cm 1cm 1.2cm, width=1.00\textwidth]{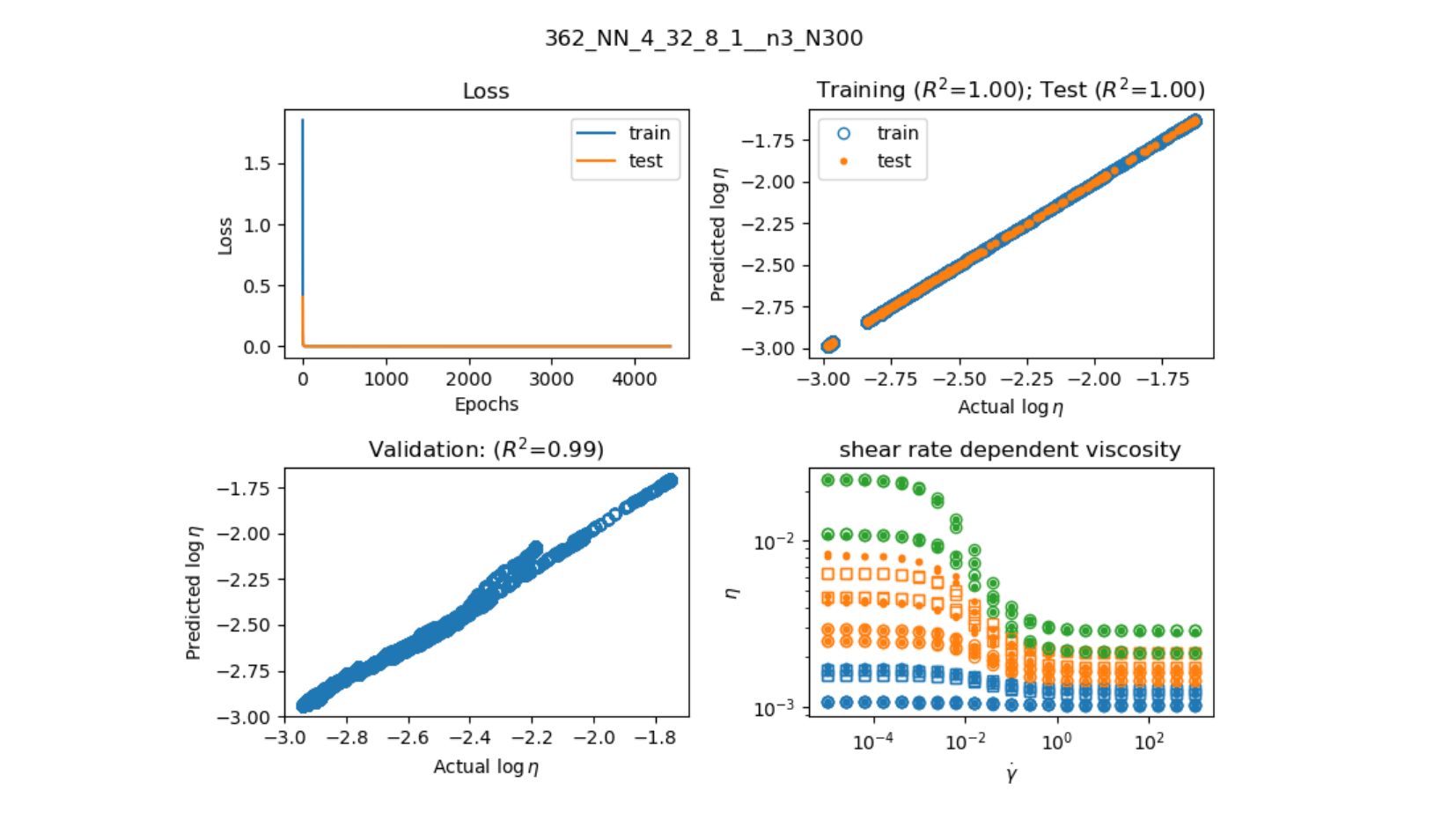}
    \caption{Results of the NN 4-32-8-1 neural network (see figure~\ref{fig:architect0_1}). The NN was trained and tested using a data set generated from $n=3$ different $\phi_p$ and for each $\phi_p$ $N=300$ different $\dot\gamma$. Top-left: loss as a function of number of epochs; Top-right predicted versus actual values of the training and test set; Bottom-left: predicted versus actual values of the validation set with seen (blue) and unseen (orange) $\phi_p$;  Bottom-right: examples of actual shear rate dependent viscosity ($\circ$ (seen $\phi_p$) and $\square$ (unseen $\phi_p$)) and predicted shear rate dependent viscosity ($\bullet$). The colors indicate different $\phi_p$.}
    \label{fig:3}
\end{figure}

Figure~\ref{fig:3} shows an example of the results of a NN for case 2, with four input nodes ($\phi_p$, $\log\dot\gamma$, $p_1$, and $p_2$), two hidden layers with 32 and 8 neurons, respectively, and one output layer $\log\eta$. This NN was trained and tested using a data set generated from $n=3$ different $\phi_p$ and for each $\phi_p$ $N=300$ different $\dot\gamma$. The top-left panel shows the loss function as a function of the number of epochs. The top-right panel shows the predicted output as a function of the actual (ground truth) output, for the training set as well as the test set. 
The predicted values and actual values of this set are plotted against each other in the left-bottom panel. 
In addition, predicted and actual curves with different $\phi_p$, seen, as well as unseen during training, as function of $\dot\gamma$ are shown in the right-bottom panel.

\begin{figure}[htbp]
    \centering
        \includegraphics[clip, trim=1cm 0.5cm 1cm 1.4cm, width=1.00\textwidth]{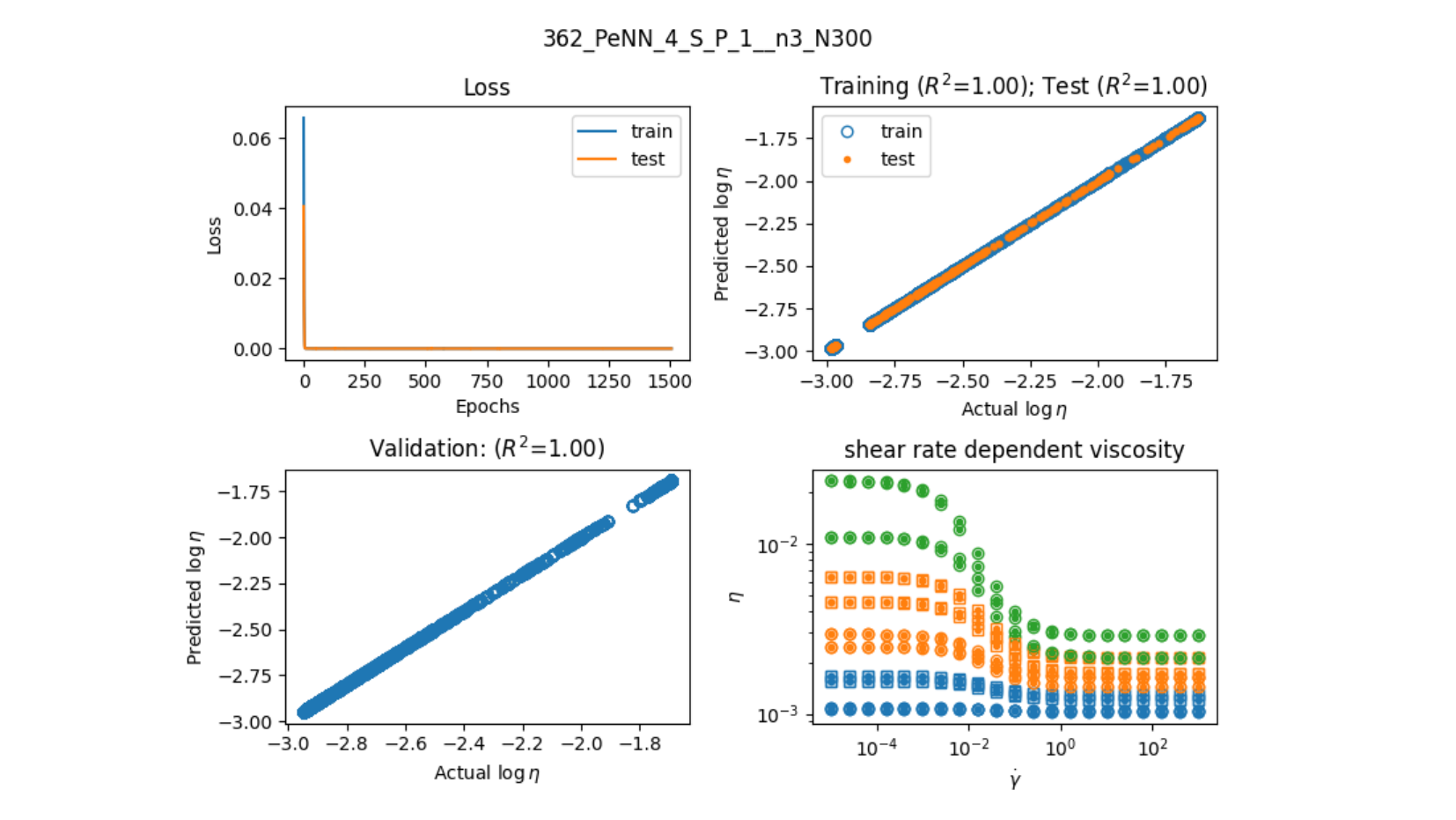}
    \caption{Results of the PeNN 4-6-2-1-1-1 physical-encoded neural network (see figure~\ref{fig:architect0_1}). The PeNN was trained and tested using a data set generated from $n=3$ different $\phi_p$ and for each $\phi_p$ $N=300$ different $\dot\gamma$.Top-left: loss as a function of number of epochs; Top-right predicted versus actual values of the training and test set; Bottom-left: predicted versus actual values of the validation set with seen (blue) and unseen (orange) $\phi_p$; Bottom-right: examples of actual ($\circ$) and predicted ($\bullet$) flow curves (colors indicate $\phi_p$.}
    \label{fig:4}
\end{figure}

Figure~\ref{fig:4} shows an example of the results of the PeNN for case 2, with four input nodes ($\phi_p$, $\log\dot\gamma$, $p_1$, and $p_2$), 2 dense NN-layers and 3 layers completely physics-encoded, of which the last one is the output layer $\log\eta$. The top-left panel shows the loss function as a function of the number of epochs. The top-right panel shows the predicted output as a function of the actual (ground truth) output, for the training set as well as the test set. 
The predicted values and actual values of this set are plotted against each other in the left-bottom panel. 
In addition, the right-bottom panel of the figure shows the predicted and ground truth viscosity as function of shear rate and for different $\phi_p$, seen, as well as unseen during training.

We conclude also for this case 2 that replacing neural nodes with physics does improve the NNs significantly. The viscosity as function of shear rate and primary particle volume fraction are predicted very well, also for values of input parameters which the model has not been trained for. Thus interpolation and extrapolation are captured well for the PeNNs.

\begin{figure}[htbp]
    \centering
        \includegraphics[clip, trim=1cm 3cm 1cm 3cm, width=1.00\textwidth]{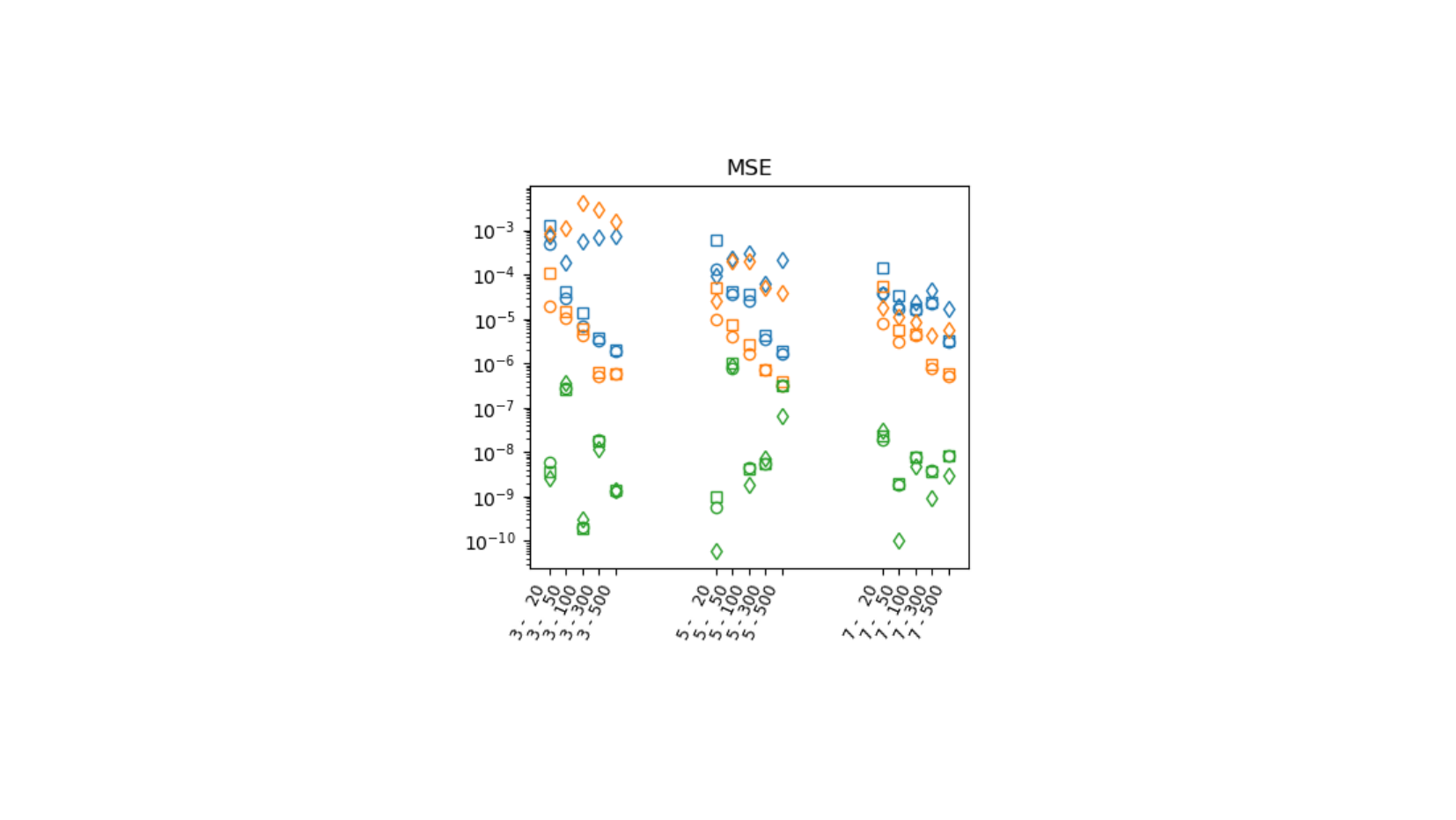}
    \caption{Mean squared error (mse) of the different NN's and PeNN for different data sets. The different colors correspond to NN 4-32-8-1 (blue), NN 4-128-32-8-1 (orange) and PeNN 4-6-2-1-1-1 (green). The different symbols correspond to the training set ($\circ$), test set ($\square$), and validation set ($\diamondsuit$). The horizontal axis correspond to different data sets generated with $n\in [3,5,7]$ different $\phi_p$'s and for each $\phi_p$, $N\in [20, 50, 100, 300, 500]$ different $\log\dot\gamma$.}
    \label{fig:5}
\end{figure}

Similarly as for case 1, the NNs can predict the viscosity curves very well, when the input parameters are part of the training set. However predictions are significantly worse for parameters not in the training set. Thus NN does not well capture interpolation and extrapolation. This is also nicely illustrated in figure~\ref{fig:5}, showing the mean squared error (mse) for the NNs and PeNN for different training, test, and validation sets. The mse is defined as ${\rm mse} = 1/m\sum_i^m (y_i-Y_i)^2$ with $Y_i$ and $y_i$ the ground truth and predicted values, respectively, and $m$ the number of data points in the set. The figure shows that the mse of the PeNN is always smaller than that of the NNs, in the order of a factor of thousand. Furthermore, it is seen that the NN's perform worse for the validation set, especially for the smaller training sets ($n=3$ and $n=5$), indicating that the NNs do not capture interpolation well. This is not seen for the PeNN, that capture interpolation well. In addition, as expected, the NNs show a decrease in mse for the training and test sets with larger data sets. Also, it is seen that the NN with 3 hidden layers performs better than the NN with 2 hidden layers. We note that the issue of optimization of the NN structures and the NN structure in the PeNN is not addressed in our study. Finally, it is observed that the mse of the PeNN seems randomly distributed between $10^{-6}$ and $10^{-10}$ for the different sizes of the training and test sets, and that the PeNN needs a much smaller data set size than the NNs to achieve a similar mse.

\section{Conclusions  and perspective}
We have illustrated for an example of a complex non-linear problem within agri-food how physics information can enrich a neural network and improve the predictability and robustness of that NN model. 

Our work uses essential physics to improve a NN model. The incorporation of physics into the architecture of the NN is such that the resulting architecture of the NN has the same architecture of a physics-based hierarchical architecture. It is this architectural hierarchical feature that is important in realizing our so-called physics-encoded NN. The term physics-encoded has some specific implications, which are clearly different from the features of other types of hybrid NNs by Faroughi et al.~\cite{faroughi2022physics}. Advantages of the physics-encoded NNs above other forms of neural networks that are enriched with physics information are the efficiency of algorithms in finite dimension settings, robustness against data scarcity, and their modular transferability into other areas. Similarly, neural-mechanistic hybrid approaches have been utilized in genome-scale metabolic models recently by Faure \etal ~\cite{faure2023neural}.

In a review by Alber \etal ~\cite{alber2019integrating}, the challenges and possible routes forward to model spatial-temporal evolution are clearly addressed. To capture the spatio-temporal dynamics, one can use partial differential equations to express physics-based conservation laws, where such conservation laws can be constructed from constitutive laws that represent the local behaviour, in combination with using ordinary differential equations and their spatial derivatives. This has been illustrated in more detail for rheology-informed neural networks by Mahmoudabadbozchelou and Jamali~\cite{mahmoudabadbozchelou2021rheology}. Interestingly, Sadaat \etal~\cite{saadat2022data,saadat2023rheologist} report the use of a platform of possible constitutive models to have the NN pick from in order to optimize its predictions for more complex fluid behavior. Interestingly, parts of constitutive models can be captured by means of applying scattering under flow, as demonstrated by Young \etal while using Small Angle X ray scattering on dilute rod suspensions~\cite{young2023scattering}. In trying to take into account spatio-temporal non-linear features, Dabri \etal~\cite{dabiri2023fractional} report the use of fractional derivatives to incorporate into the NN models. Fractional derivatives are known to represent the presence of memory effects, which may be uncovered by introducing hidden variables that describe local effects, as addressed in~\cite{weinan2020integrating}. In that work, an interesting example for introducing temporal (dynamical) information has been addressed in the form of so-called “recurrent neural networks”. The neural networks are machine learning models for time series. These models use hidden variables, making the relationships, as expressed in the models, local.  If no hidden variables are being used, one effectively introduces memory effects.

The physics-encoded NNs allows continued learning as opposed to instance learning. Instance learning implies that it is required to retrain an entire network for predicting outcomes in a new setting for that network. Continued learning resembles a feature of intelligence. In this view, physics-encoded neural networks should not be viewed as an “AI” model but an “I” model, i.e. leaving out the term “Artificial”. In respect to learning, it is noteworthy that recently Zador~\cite{zador2019critique} has argued that one may distinguish several levels of learning within animals, as opposed to the technical usage of the term learning within NN’s. Within animals, there exist learned and innate mechanisms in executing functions. The innate mechanisms are encoded in the genome, which provides the rules that wire up the brain, for the behavioral programs for many functions (walking, swimming, etc.). The wiring of the brain is not explicitly programmed but will evolve during development, on the basis of a set of rules given by the coding. Interestingly, ANNs have to be optimized according to what is learned during their evolution and the learning during their functioning. In contrast, animals learn only in their functioning, as the innate part has been encoded. In analogy to the above, in physics-encoded NNs, the physics provides the set of rules, that allow for the explicit efficient wiring of the NNs during their training period. In this sense, one always should use NNs that build on previous solutions (use their learning). This represents a “real” learning phase where information is stored in a structured manner, suitable for building new information on top of that information, in a congruent manner, instead of building a structure all the time on the basis of data available, en re-iterating this latter learning. In fact, the physical information encoding implies a form of “real” intelligence. This explains the advantage of using physics to encode the NN wiring. This wiring is not random but instead is containing information, obtained from exposure to its surroundings and subsequently storing that information which, if used again, explicates learning. This view is extended into a formal theory of clever computing of systems as expressed by Jaeger \etal ~\cite{jaeger2023toward}, where the ideal computing of systems occurs as a bottom-up activity that structures the processes along which the computing takes place (cybernetic), using physically observables (physics-encoded), instead of the classic computing systems that describe the processing along structures that are present (algorithmic). 

In summary, our work is straightforward and uses essential physics understanding to improve a NN model. The NNs can predict the viscosity curves very well, when the input parameters are part of the training set. Predictions are significantly worse for parameters not in the training set. Thus NN does not sufficiently capture interpolation and extrapolation. The physics model may not be entirely covering all aspects, but the physics will be directing the number of possibilities for the NN while optimizing during its learning stage. The incorporation of the physics into the architecture of the NN such that the resulting architecture of the NN has the same architecture of the physics-based hierarchical architecture. It is this architectural characteristic that is important in being embedded into the NN, which is different from what has been reported until now. 

Using physics-based information introduces information in NN’s that has an experimental basis, and the information added incorporates the existing and known natural sequence of events and their hierarchy, i.e. it incorporates the way that nature shows itself to us.
In the same spirit, one can imagine to make multi-scale hybrid models with neural networks encoded by general accepted relations (other than physics-based relations). Such hybrid models may be relating physical properties of a food with its sensory perception and health impact, or relating crop properties with (local) climate circumstances. Using a similar hybrid model relating physical properties of foods with crop properties will allow us to relate climate to food properties with sensory perception and health impact. These relations represent an important part of the agri-food system. We note that the modules may be sequential and/or nested. Having these relations available in the form of a modular hybrid model will make the agri-food system more robust and adaptable, to climate changes. 

\section*{Acknowledgement}
We acknowledge funding from the Dutch "Sectorplan Techniek II".


\begin{thebibliography}{10}

\bibitem{van2014complex}
H.~G. van Mil, E.~Foegeding, E.~J. Windhab, N.~Perrot, and E.~Van Der~Linden,
  ``A complex system approach to address world challenges in food and
  agriculture,'' {\em Trends in food science \& technology}, vol.~40, no.~1,
  pp.~20--32, 2014.

\bibitem{perrot2016some}
N.~Perrot, H.~De~Vries, E.~Lutton, H.~G. Van~Mil, M.~Donner, A.~Tonda,
  S.~Martin, I.~Alvarez, P.~Bourgine, E.~Van Der~Linden, {\em et~al.}, ``Some
  remarks on computational approaches towards sustainable complex agri-food
  systems,'' {\em Trends in Food Science \& Technology}, vol.~48, pp.~88--101,
  2016.

\bibitem{weinan2020integrating}
E.~Weinan, J.~Han, L.~Zhang, {\em et~al.}, ``Integrating machine learning with
  physics-based modeling,'' {\em arXiv preprint arXiv:2006.02619}, 2020.

\bibitem{jia2021physics}
X.~Jia, J.~Willard, A.~Karpatne, J.~S. Read, J.~A. Zwart, M.~Steinbach, and
  V.~Kumar, ``Physics-guided machine learning for scientific discovery: An
  application in simulating lake temperature profiles,'' {\em ACM/IMS
  Transactions on Data Science}, vol.~2, no.~3, pp.~1--26, 2021.

\bibitem{daw2017physics}
A.~Daw, A.~Karpatne, W.~Watkins, J.~Read, and V.~Kumar, ``Physics-guided neural
  networks (pgnn): An application in lake temperature modeling,'' {\em arXiv
  preprint arXiv:1710.11431}, 2017.

\bibitem{read2019process}
J.~S. Read, X.~Jia, J.~Willard, A.~P. Appling, J.~A. Zwart, S.~K. Oliver,
  A.~Karpatne, G.~J. Hansen, P.~C. Hanson, W.~Watkins, {\em et~al.},
  ``Process-guided deep learning predictions of lake water temperature,'' {\em
  Water Resources Research}, vol.~55, no.~11, pp.~9173--9190, 2019.

\bibitem{hanson2020predicting}
P.~C. Hanson, A.~B. Stillman, X.~Jia, A.~Karpatne, H.~A. Dugan, C.~C. Carey,
  J.~Stachelek, N.~K. Ward, Y.~Zhang, J.~S. Read, {\em et~al.}, ``Predicting
  lake surface water phosphorus dynamics using process-guided machine
  learning,'' {\em Ecological Modelling}, vol.~430, p.~109136, 2020.

\bibitem{faroughi2022physics}
S.~A. Faroughi, N.~Pawar, C.~Fernandes, S.~Das, N.~K. Kalantari, and S.~K.
  Mahjour, ``Physics-guided, physics-informed, and physics-encoded neural
  networks in scientific computing,'' {\em arXiv preprint arXiv:2211.07377},
  2022.

\bibitem{alber2019integrating}
M.~Alber, A.~Buganza~Tepole, W.~R. Cannon, S.~De, S.~Dura-Bernal,
  K.~Garikipati, G.~Karniadakis, W.~W. Lytton, P.~Perdikaris, L.~Petzold, {\em
  et~al.}, ``Integrating machine learning and multiscale
  modeling—perspectives, challenges, and opportunities in the biological,
  biomedical, and behavioral sciences,'' {\em NPJ digital medicine}, vol.~2,
  no.~1, p.~115, 2019.

\bibitem{mahmoudabadbozchelou2021rheology}
M.~Mahmoudabadbozchelou and S.~Jamali, ``Rheology-informed neural networks
  (rhinns) for forward and inverse metamodelling of complex fluids,'' {\em
  Scientific reports}, vol.~11, no.~1, p.~12015, 2021.

\bibitem{saadat2022data}
M.~Saadat, M.~Mahmoudabadbozchelou, and S.~Jamali, ``Data-driven selection of
  constitutive models via rheology-informed neural networks (rhinns),'' {\em
  Rheologica Acta}, vol.~61, no.~10, pp.~721--732, 2022.

\bibitem{young2023scattering}
C.~D. Young, P.~T. Corona, A.~Datta, M.~E. Helgeson, and M.~D. Graham,
  ``Scattering-informed microstructure prediction during lagrangian evolution
  (simple)--a data-driven framework for modeling complex fluids in flow,'' {\em
  arXiv preprint arXiv:2305.03792}, 2023.

\bibitem{dabiri2023fractional}
D.~Dabiri, M.~Saadat, D.~Mangal, and S.~Jamali, ``Fractional rheology-informed
  neural networks for data-driven identification of viscoelastic constitutive
  models,'' {\em Rheologica Acta}, pp.~1--12, 2023.

\bibitem{willard2020}
J.~Willard, X.~Jia, M.~Steinbach, V.~Kumar, and S.~Xu, ``Integrating
  physics-based modeling with machine learning: A survey,'' {\em arXiv preprint
  arXiv:2003.04919}, vol.~1, p.~34, 2020.

\bibitem{takeishi2021}
N.~Takeishi and A.~Kalousis, ``Physics-integrated variational autoencoders for
  robust and interpretable generative modeling,'' {\em Advances in Neural
  Information Processing Systems}, vol.~34, pp.~14809--14821, 2021.

\bibitem{peng2020multiscale}
G.~C.~Y. Peng, M.~Alber, A.~B. Tepole, W.~Cannon, S.~De, S.~Dura-Bernal,
  K.~Garikipati, G.~Karniadakis, W.~W. Lytton, P.~Perdikaris, L.~Petzold, and
  E.~Kuhl, ``Multiscale modeling meets machine learning: What can we learn?,''
  {\em Archives of Computational Methods in Engineering}, vol.~28,
  pp.~1017--1037, 2021.

\bibitem{lie2023machine}
A.~Lie-Piang, A.~Garre, T.~Nissink, N.~van Beek, A.~van~der Padt, and R.~Boom,
  ``Machine learning to quantify techno-functional properties-a case study for
  gel stiffness with pea ingredients,'' {\em Innovative Food Science \&
  Emerging Technologies}, vol.~83, p.~103242, 2023.

\bibitem{bhagya2022comprehensive}
G.~Bhagya~Raj and K.~K. Dash, ``Comprehensive study on applications of
  artificial neural network in food process modeling,'' {\em Critical reviews
  in food science and nutrition}, vol.~62, no.~10, pp.~2756--2783, 2022.

\bibitem{datta2022computer}
A.~Datta, B.~Nicola{\"\i}, O.~Vitrac, P.~Verboven, F.~Erdogdu, F.~Marra,
  F.~Sarghini, and C.~Koh, ``Computer-aided food engineering,'' {\em Nature
  Food}, pp.~1--11, 2022.

\bibitem{derkach2009rheology}
S.~R. Derkach, ``Rheology of emulsions,'' {\em Advances in colloid and
  interface science}, vol.~151, pp.~1--23, 2009.

\bibitem{rao2014rheology}
M.~A. Rao, {\em Rheology of Fluid and Semisolid Foods.pdf}.
\newblock Springer, 2014.

\bibitem{larson2019}
R.~G. Larson and Y.~Wei, ``A review of thixotropy and its rheological
  modeling,'' {\em Journal of Rheology}, vol.~63, pp.~477--501, 2019.

\bibitem{mcclements2023modeling}
D.~J. McClements, ``Modeling the rheological properties of plant-based foods:
  Soft matter physics principles,'' {\em Sustainable Food Proteins}, 2023.

\bibitem{quemada1977}
D.~Quemada, ``Rheology of concentrated disperse systems and minimum energy
  dissipation principle - i. viscosity-concentration relationship,'' {\em
  Rheologica Acta}, vol.~16, pp.~82--94, 1977.

\bibitem{quemada1998a}
D.~Quemada, ``Rheological modelling of complex fluids. i. the concept of
  effective volume fraction revisited,'' {\em The European Physical
  Journal-Applied Physics}, vol.~1, pp.~119--127, 1998.

\bibitem{berli2000rheological}
C.~L. Berli and D.~Quemada, ``Rheological modeling of microgel suspensions
  involving solid-liquid transition,'' {\em Langmuir}, vol.~16, pp.~7968--7974,
  2000.

\bibitem{berli2002modeling}
C.~L. Berli, D.~Quemada, and A.~Parker, ``Modelling the viscosity of depletion
  flocculated emulsions,'' {\em Colloids and Surfaces A: Physicochemical and
  Engineering Aspects}, vol.~203, pp.~11--20, 4 2002.

\bibitem{quemada2002energy}
D.~Quemada and C.~Berli, ``Energy of interaction in colloids and its
  implications in rheological modeling,'' {\em Advances in colloid and
  interface science}, vol.~98, pp.~51--85, 2002.

\bibitem{jansen2001viscosity}
K.~M.~B. Jansen, W.~G.~M. Agterof, and J.~Mellema, ``Viscosity of surfactant
  stabilized emulsions,'' {\em Citation: Journal of Rheology}, vol.~45,
  p.~1359, 2001.

\bibitem{faure2023neural}
L.~Faure, B.~Mollet, W.~Liebermeister, and J.-L. Faulon, ``A neural-mechanistic
  hybrid approach improving the predictive power of genome-scale metabolic
  models,'' {\em Nature Communications}, vol.~14, no.~1, p.~4669, 2023.

\bibitem{saadat2023rheologist}
M.~Saadat, D.~Mangal, and S.~Jamali, ``A rheologist's guideline to data-driven
  recovery of complex fluids' parameters from constitutive models,'' {\em
  Digital Discovery}, 2023.

\bibitem{zador2019critique}
A.~M. Zador, ``A critique of pure learning and what artificial neural networks
  can learn from animal brains,'' {\em Nature communications}, vol.~10, no.~1,
  p.~3770, 2019.

\bibitem{jaeger2023toward}
H.~Jaeger, B.~Noheda, and W.~G. Van Der~Wiel, ``Toward a formal theory for
  computing machines made out of whatever physics offers,'' {\em Nature
  Communications}, vol.~14, no.~1, p.~4911, 2023.

\end{thebibliography}

\end{document}